\documentclass[a4paper,10pt,twoside]{cpc-hepnp}

\usepackage{multicol}
\usepackage{graphicx}
\usepackage{booktabs}
\usepackage{amssymb,bm,mathrsfs,bbm,amscd}
\usepackage[tbtags]{amsmath}
\usepackage{lastpage}
\usepackage{CJK}
\usepackage{multirow}
\usepackage{subfigure}

\begin{document}
\begin{CJK*}{GBK}{song}

\fancyhead[c]{\small Chinese Physics C~~~Vol. 37, No. 1 (2013)
010201} \fancyfoot[C]{\small 010201-\thepage}

\footnotetext[0]{Received 14 March 2009}

\title{Study of digital pulse shape discrimination method for n-$\gamma$ separation
of EJ-301 liquid scintillation detector\thanks{Supported by National Natural Science Foundation of China (91226107, 11305229) and the Strategic Priority Research Program of the Chinese Academy of Sciences (XDA03030300 ) }}

\author{%
      WAN Bo$^{1,2}$
\quad ZHANG Xue-Ying$^{1;1)}$\email{zhxy@impcas.ac.cn}%
\quad CHEN Liang$^{1}$
\quad GE Hong-Lin$^{1}$\\
\quad MA Fei$^{1}$
\quad ZHANG Hong-Bin$^{1}$
\quad JU Yong-Qin$^{1}$\\
\quad ZHANG Yan-Bin$^{1}$
\quad Li Yan-Yan$^{1,2}$
\quad XU Xiao-Wei$^{1,2}$
}
\maketitle

\address{%
$^1$ Institute of Modern Physics, Chinese Academy of Sciences, Lanzhou 730000, China\\
$^2$ University of Chinese Academy of Sciences, Beijing 100049, China\\
 }

\begin{abstract}
A digital pulse shape discrimination system based on a programmable module NI-5772 has been established and tested with EJ-301 liquid scintillation detector.  The module was operated by means of running programs developed in LabVIEW with the sampling frequency up to 1.6GS/s. Standard gamma sources $^{22}$Na, $^{137}$Cs and $^{60}$Co were used to calibrate the EJ-301 liquid scintillation detector, and the gamma response function has been obtained. Digital algorithms for charge comparison method and zero-crossing method have been developed. The experimental results showed that both digital signal processing (DSP) algorithms could discriminate neutrons from gamma-rays. Moreover, the zero-crossing method shows better n-$\gamma$ discrimination at 80 keVee and lower, whereas the charge comparison method gives better results at higher thresholds. In addition, the figure-of-merit (FOM) of two different dimension detectors were extracted at 9 energy thresholds, and it was found that the smaller one presented a better n-$\gamma$ separation property for fission neutrons.
\end{abstract}



\footnotetext[0]{\hspace*{-3mm}\raisebox{0.3ex}{$\scriptstyle\copyright$}2013
Chinese Physical Society and the Institute of High Energy Physics
of the Chinese Academy of Sciences and the Institute
of Modern Physics of the Chinese Academy of Sciences and IOP Publishing Ltd}%

\begin{multicols}{2}

\section{Introduction}

For the last decades, concepts of accelerator driving subcritical systems (ADS) and spallation neutron sources (SNS) have been proposed and constructed based on the spallation reactions~\cite{lab1,lab2,lab3}. High intensity neutron flux and gamma-rays would be produced around the spallation targets. Kinds of neutron detectors have been used to monitor the leakage neutrons in real time in mixed neutrons-gamma-rays radiation environments, such as fission chamber, $^{3}$He counter and organic liquid scintillation detector et al. Among of these detectors, the EJ-301 liquid scintillation detectors have been widely employed because of their excellent neutron-gamma discrimination, high efficiency for fast neutron detection and the superior time resolution. Aiming to discriminate neutrons from gamma-rays for the EJ-301 liquid scintillation detector, two most common pulse shape discrimination (PSD) methods could be used. One of them is the zero-crossing method, which extracts the zero-crossing time of the doubly differential bipolar pulse~\cite{lab4,lab5}. Another is the charge comparison method based on independent measurements of the integrated charge over two different time regions of the pulse~\cite{lab6,lab7}.

Recently, with the development of the field programmable gate array (FPGA) technology and computer CPUs, the digital signal processing (DSP) is now possible. The major difference between analog and DSP techniques is that with the digital method, the current pulse from the anode of photomultiplier tube (PMT) is digitized immediately and all operations are carried out in software package. For the analog PSD methods, the operations have to be implemented based on CAMAC or VME modules along with a series of complicated analog circuits. The DSP system offers significant advantages over the analog system in the area of conveniences, real-time properties and cramped construction by eliminating extra electronic modules. Moreover, prior to this work, some research groups have investigated the DSP method, and performed the comparisons of the n-$\gamma$ separation results using the zero-crossing method and charge comparison method, respectively~\cite{lab8,lab9}. However, their experimental results showed different behaviours on the ability of n-$\gamma$ discrimination. For example, when comparing the zero-crossing method and charge comparison method, results in Ref.~\cite{lab8} suggested the former showed better n-$\gamma$ separation property, whereas the conclusion in Ref.~\cite{lab9} was contrary. Therefore, it is very essential to develop DSP method, and timely to reinvestigate the zero-crossing method and charge comparison method.

In this paper, a digital acquisition system based on the NI-5772 adapter module has been developed. The algorithms of digital pulse shape discrimination method for n-$\gamma$ separation of EJ-301 liquid scintillation detector were described in Section 2. The liquid scintillation detector was calibrated by the standard gamma-rays sources, and the capacity of n-$\gamma$ discrimination for the new system was tested with $^{252}$Cf neutron source. At the same time, the capacity of n-$\gamma$ discrimination for two different dimension liquid scintillation detectors was also studied.

\section{Experimental details}

\subsection{Algorithm for PSD}

Fig.~\ref{fig1}(a) presents the typical current pulses related to neutrons and gamma-rays. Pulse generated by gamma-ray decays faster to the baseline than the neutron induced pulse. The major difference between these two pulses occurs in their tail. It is this fact that neutrons and gamma-rays could be discriminated by analyzing the contributions of slow components to the total light output.

In this paper, two types of digital data processing techniques are employed for n-$\gamma$ discrimination. One of them is the digital implementation of the conventional charge comparison method. Q$_{total}$ and Q$_{slow}$ represent the total charge of the current pulse and the charge in the slow components, respectively, as shown in Fig.~\ref{fig1}(a). Thus neutrons and gamma-rays could be separated accurately in mixed radiation fields through comparing the difference of Q$_{slow}$ for each pulse. The other algorithm is zero-crossing method. In Fig.~\ref{fig1}(b), the bipolar pulse is obtained by analyzing a digital differentiator-integrator-integrator (C$_1$R$_1$-(R${_2}$C${_2}$)$^2$) algorithm to the PMT current signal. Therefore, the difference in the tail of different PMT signals is reflected in the zero-crossing time of the bipolar pulses. To exploit the difference in the zero-crossing time as a parameter for n-$\gamma$ discrimination, a digital algorithm acting as a constant fraction discriminator (CFD) is used to determine the zero-crossing points.

\begin{center}

\includegraphics[width=7cm]{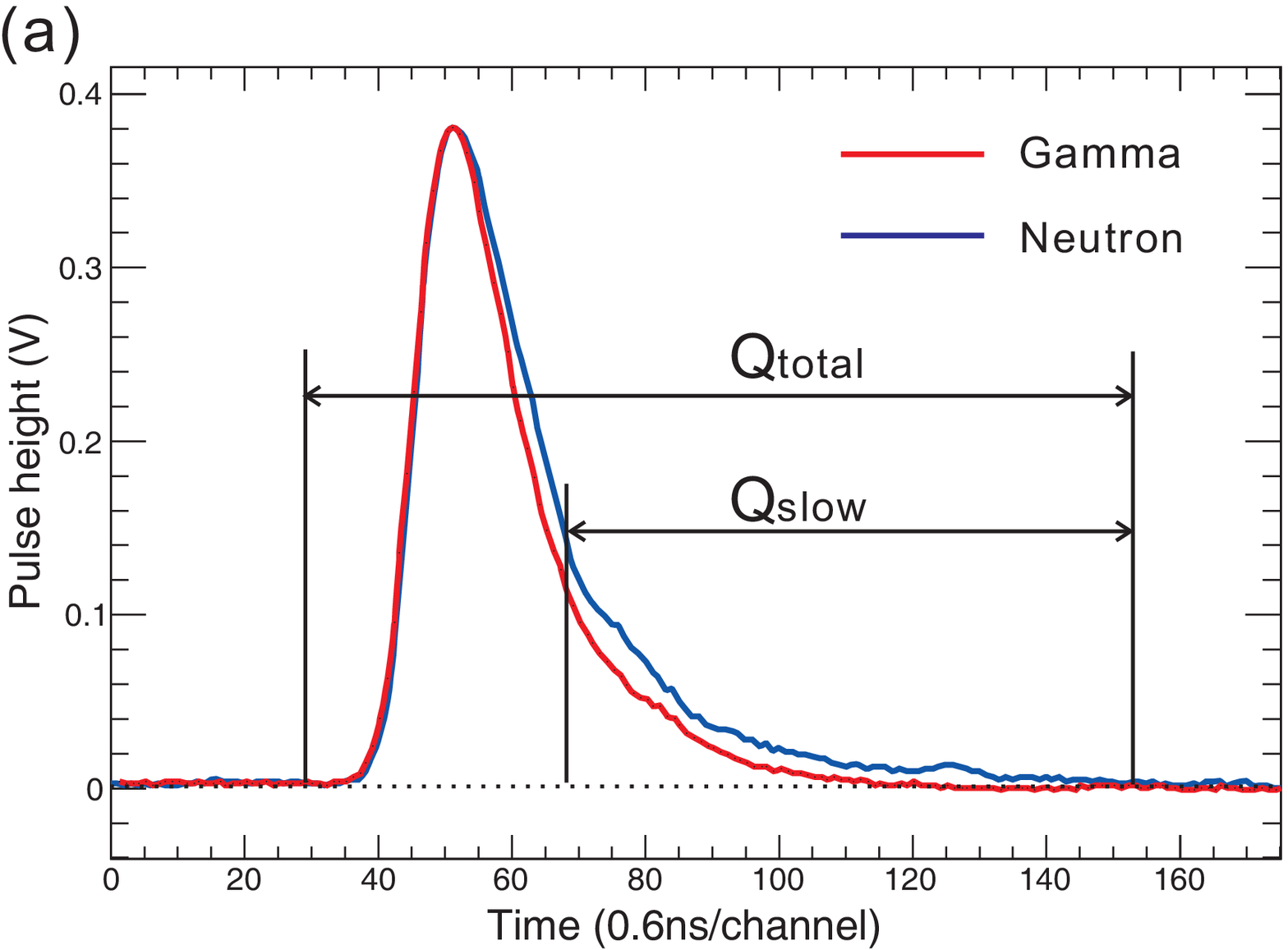}\quad
\includegraphics[width=8cm]{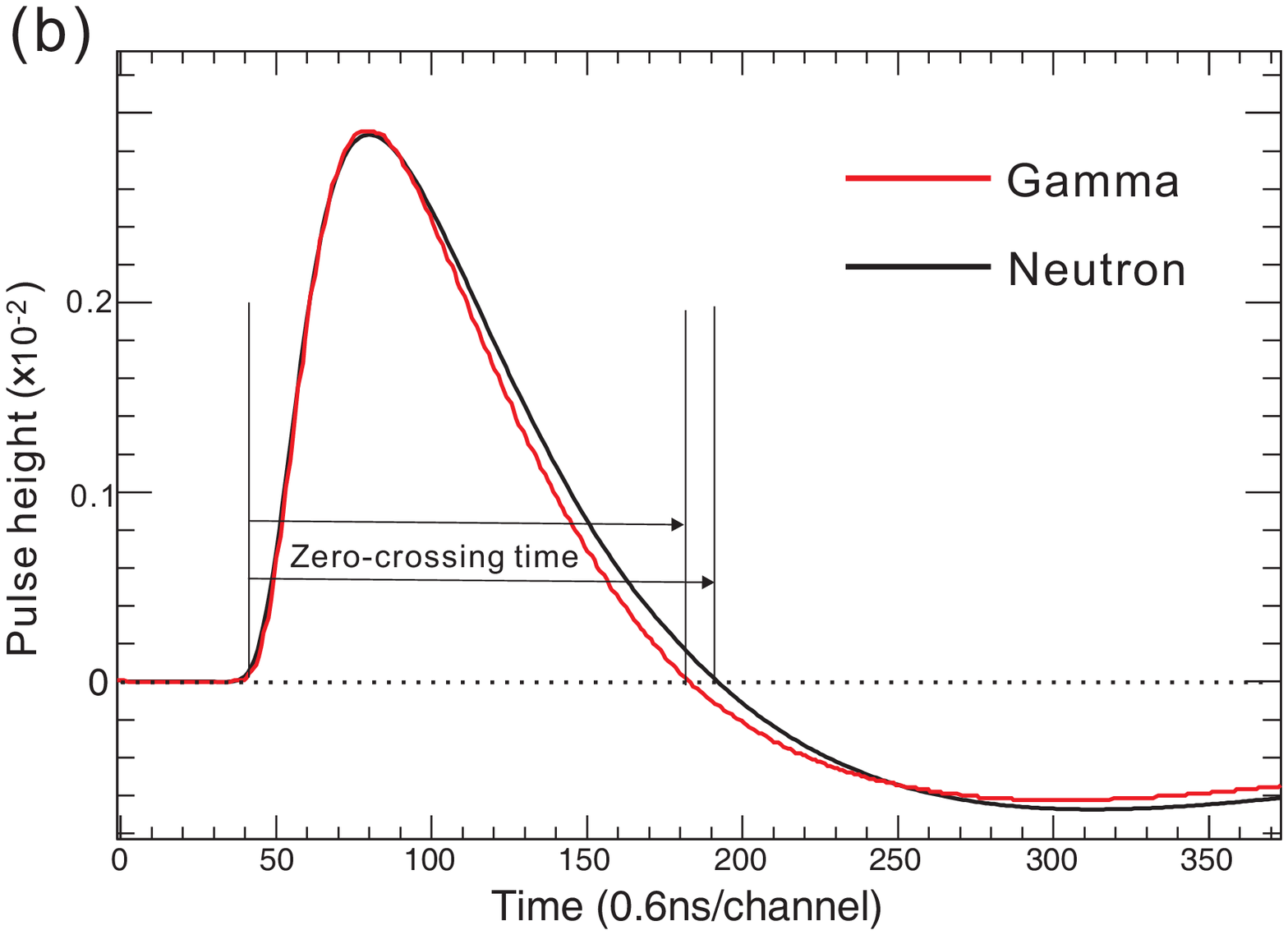}
\figcaption{ (a) Typical PMT signals induced by neutrons and gamma-rays. The gamma-ray pulse decays more quickly than the neutron pulse, thus small difference can be seen in the tail. (b) PMT signals after the pulse shaping process. The $\gamma$-ray and neutron pulses cross the zero line at different times.}
\label{fig1}
\end{center}

\subsection{Experiment setup}

\begin{center}
\includegraphics[width=8cm]{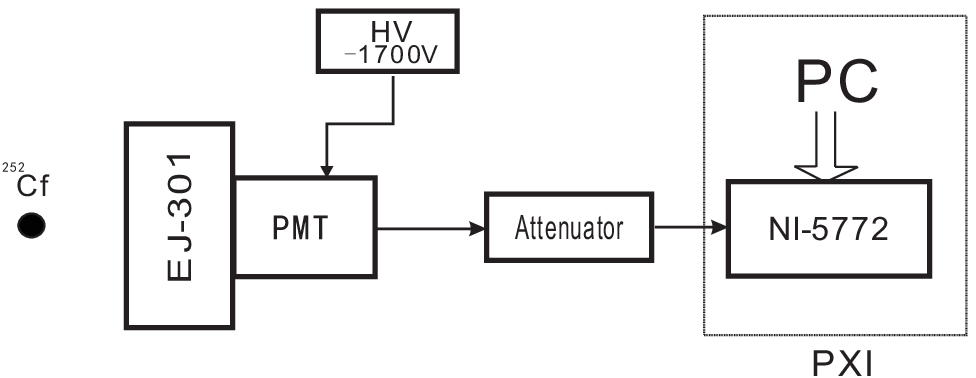}
\figcaption{\label{fig2}   The diagram of experiment setup. }
\end{center}

The diagram of experimental setup and electronics are shown in Fig.~\ref{fig2}.
A cylindrical EJ-301 liquid scintillation detector of 2-inch in diameter and 2-inch long coupled to a ET 9813KB PMT was used to perform the measurements. The detector was operated with a negative voltage of 1700V. The current pulses from anode of PMT were transferred to an attenuator and then directly digitized using a digital oscilloscope (NI-5772), which could digitize the waveform by running dedicated LabVIEW data acquisition package. Since this acquisition system could only digitize the signals whose maximum voltage range was between -1V and 1V, the 9dB attenuator was employed. In the present work, the digital oscilloscope worked at self-triggering mode with a sampling rate of 1.6GS/s and 12 bit resolution. A $^{252}$Cf neutron source with the intensity of 1.0$\times$$10^4$ n/s was used to test the n-$\gamma$ discrimination capability of this system, and standard gamma sources were also used for energy calibration of EJ-301 liquid scintillation detector. Off-line data analysis was performed by running ROOT script files written in C++ language.  In addition, to compare the capability of n-$\gamma$ discrimination for scintillator in different volume, another larger liquid scintillation detector of 2-inch in diameter and 4-inch long was also employed.

\subsection{Energy calibration}

In order to know the specific relationship between light output distribution and the realistic energy deposited in scintillator, and determine the effective neutron detection thresholds, the gamma energy calibration for the EJ-301 liquid scintillation detector was done with standard gamma sources $^{22}$Na, $^{137}$Cs and $^{60}$Co. Because the elements of liquid scintillator have very low photoelectric effect cross section, the obtained spectrum are mainly due to Compton scattering electrons~\cite{lab10}. The position of the maximum of Compton scattering electrons could be accurately determined through comparing the measured light output spectrum with simulated ones using different Monte Carlo simulation packages, such as GEANT4~\cite{lab11,lab12}, FLUKA~\cite{lab13,lab14} and GRESP7~\cite{lab15}. Among of these packages, it has been proved that the GRESP7 code is very reliable for energy calibration. More details have been described in Ref.~\cite{lab16,lab17}.

In this work, the light output spectrum of the detector for $^{22}$Na, $^{137}$Cs and $^{60}$Co gamma sources were simulated by GRESP7 code.
Fig.~\ref{fig3} shows the experimental and GRESP7 simulated Compton electron spectrum of $^{137}$Cs source. As shown in Fig.~\ref{fig3}, the maximum Compton edge is corresponding to 81$\%$ of the Compton electron distribution.
The results of calibration together with the light output of $^{22}$Na, $^{137}$Cs and $^{60}$Co are shown in Fig.~\ref{fig4}, in which the experimental data is fitted by linear polynomial.

The gamma response function of EJ-301 liquid scintillation detector is expressed by:
\begin{equation}
L=4.94*E_e-0.095,
\end{equation}
where L is the light output, $E_e$ is the deposited electron energy for liquid scintillator in MeV. This calibration is implemented to calculate the equivalent electron energy (MeVee), where 1 MeVee corresponds to the total light output induced by a 1 MeV electron.

\begin{center}
\includegraphics[width=8cm]{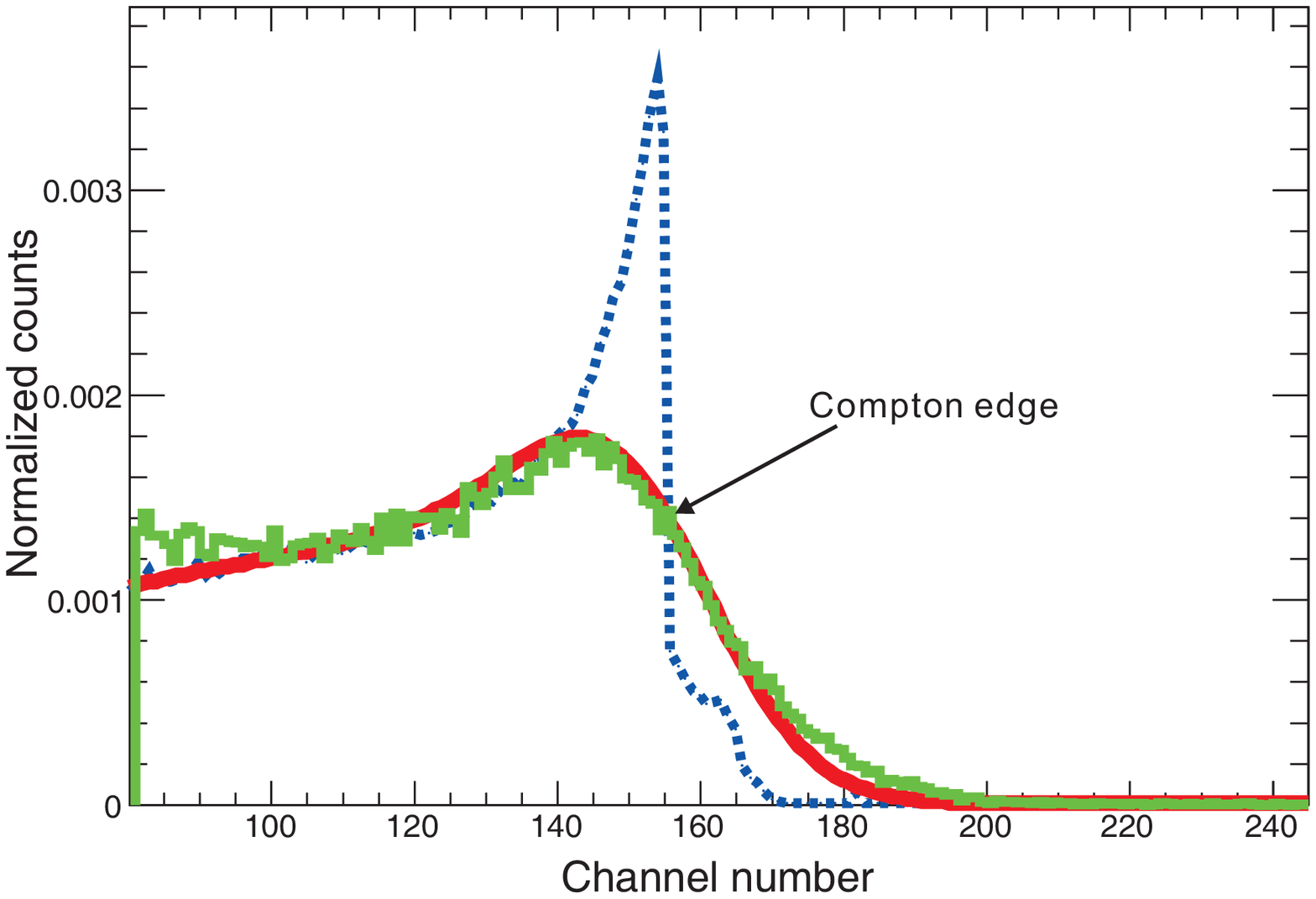}
\figcaption{\label{fig3}   Simulated Compton electron spectrum of $^{137}$Cs source (blue dash line) and the same spectrum folded with energy resolution (red solid line). Green solid line is experimental light output of $^{137}$Cs. }
\end{center}

\begin{center}
\includegraphics[width=8cm]{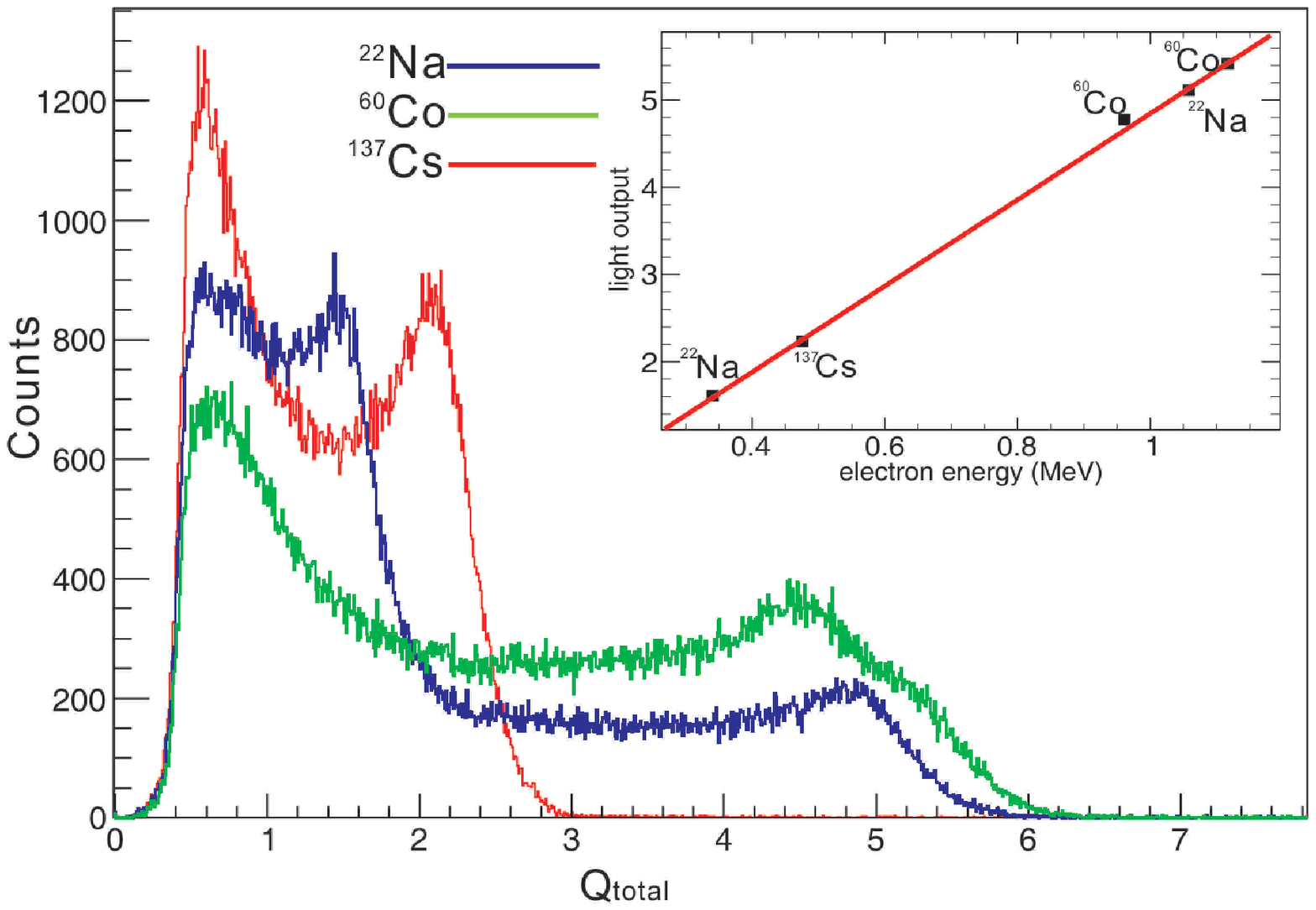}
\figcaption{\label{fig4}   Experimental light output for $^{22}$Na, $^{137}$Cs and $^{60}$Co gamma sources. The inset shows the calibration result using the Compton edges from the detected spectrum. }
\end{center}

\section{Results and discussion}

\subsection{Charge comparison method}

Two-dimensional scatter plot of the n-$\gamma$ discrimination using charge comparison method at an energy threshold of 180 keVee is shown in Fig.~\ref{fig5}.
There are two separated strips in the 2D graph where the upper one is connected with neutrons because of larger Q$_{slow}$ for equal light output. The overlap is due to the case of very low energy neutrons and gamma rays, which is the region that determines how well the neutrons are discriminated from gamma-rays.

\begin{center}
\includegraphics[width=8cm]{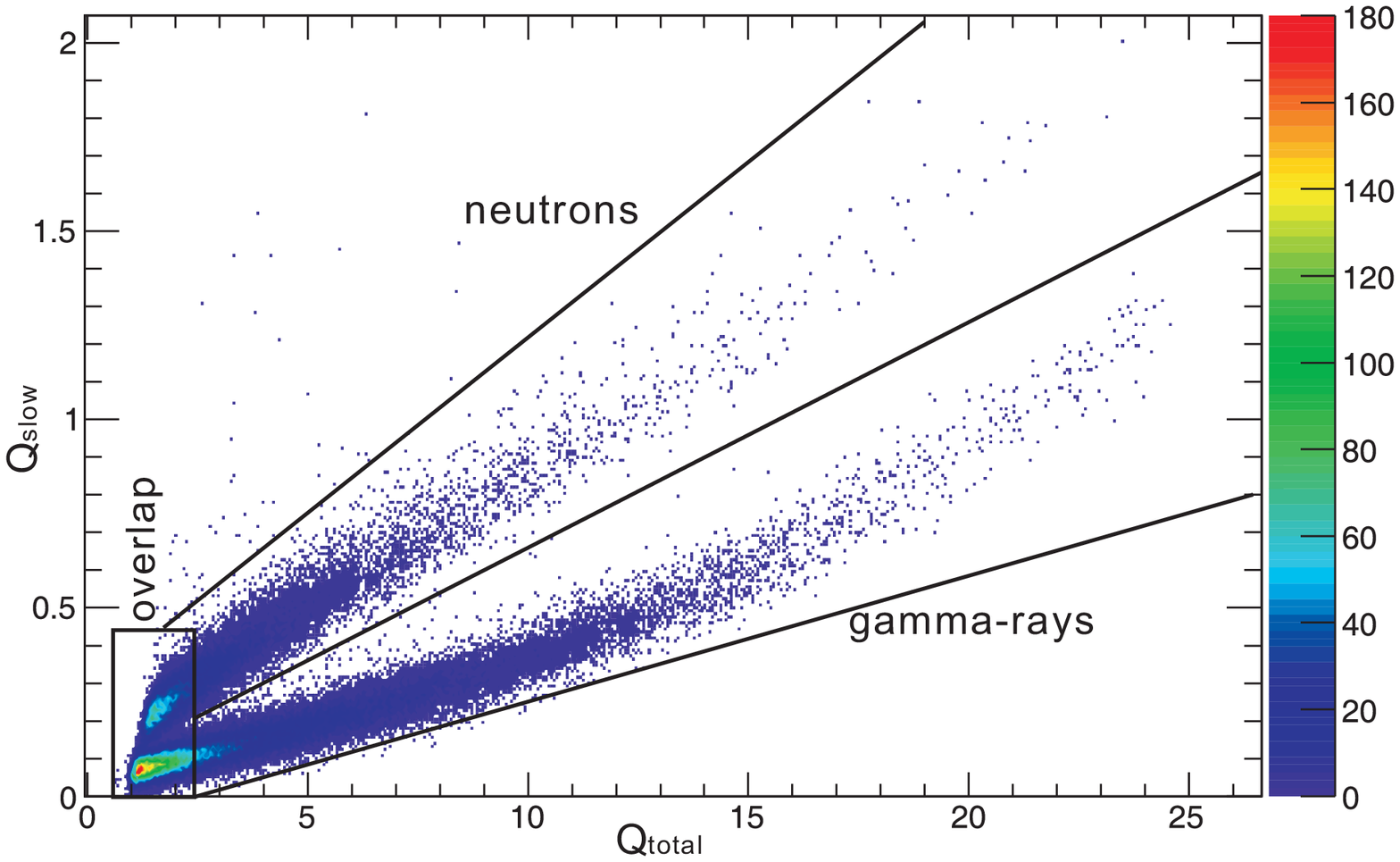}
\figcaption{\label{fig5} 2D plot of Q$_{total}$ versus Q$_{slow}$ for $^{252}$Cf source, the energy threshold is 180 keVee. }
\end{center}

The quality of n-$\gamma$ discrimination depend on several reasons, such as the employed algorithms, the sampling rate and the type of detectors and so on. In order to check the character of n-$\gamma$ separation, a figure-of-merit (FOM) is introduced:

\begin{equation}
FOM=\frac{\Delta}{\Delta \gamma +\Delta n},
\end{equation}
where $\Delta$ is the separation between the peaks of the neutron and gamma events. The $\Delta$$\gamma$ and $\Delta$n are the FWHM (full width at half maximum) of the gamma and neutron peaks, respectively. The procedure of extracting the FOM is illustrated with 1D histogram of Q$_{slow}$/Q$_{total}$, as shown in Fig.~\ref{fig6}.

\begin{center}
\includegraphics[width=8cm]{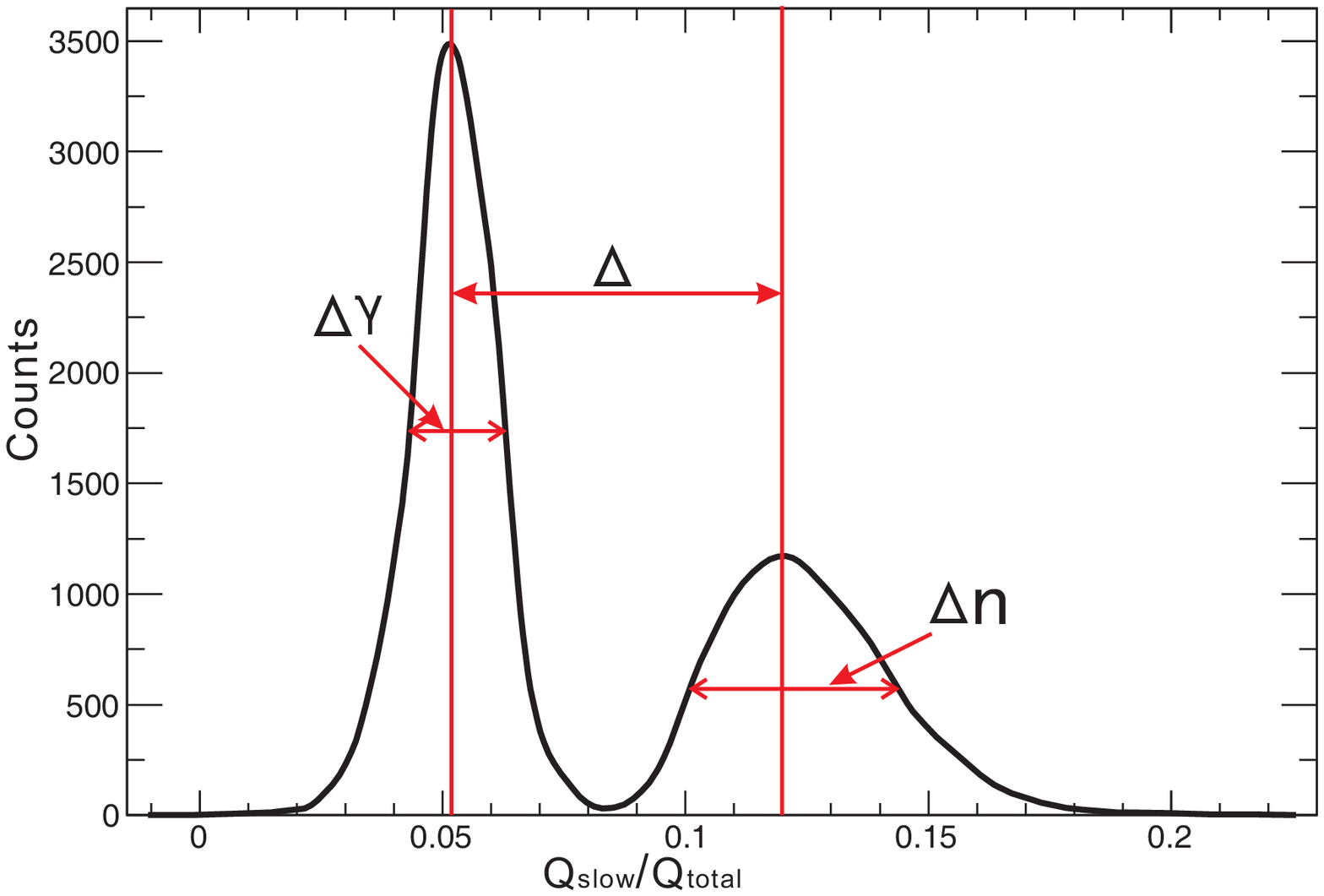}
\figcaption{\label{fig6} The procedure of calculating the FOM. }
\end{center}

FOMs over different time intervals were extracted for the purpose of evaluating the optimum integration intervals for  Q$_{slow}$ and Q$_{total}$. The integration intervals were optimized for minimizing the number of misclassified pulses. In this work, the total integration was calculated from beginning of the pulse to the 50ns after the pulse maximum for all the signals. Meanwhile, FOMs for different time intervals of Q$_{slow}$ are shown in Fig.~\ref{fig7} at 5 energy thresholds (60keVee, 120keVee, 180keVee, 240keVee and 300keVee). It is found that the optimum n-$\gamma$ separation could be derived since the tail integration started 21ns and ended 50ns from the pulse maximum for the all studied energy thresholds. The extracted optimum integration intervals will be used for charge comparison algorithm in following parts.

\begin{center}
\includegraphics[width=8cm]{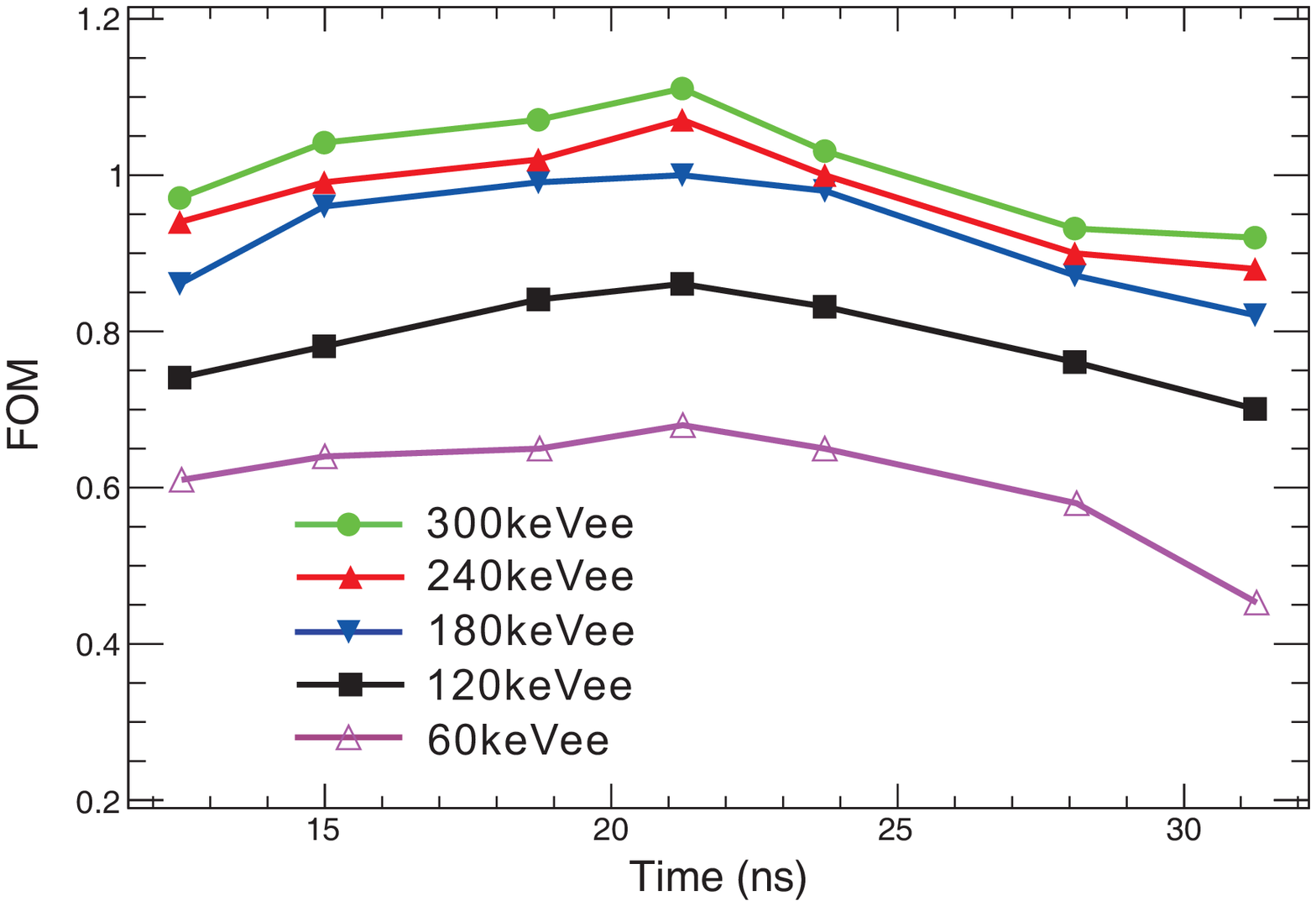}
\figcaption{\label{fig7}  FOMs for different integration intervals of Q$_{slow}$, the energy thresholds are 60keVee, 120keVee, 180keVee, 240keVee and 300keVee. }
\end{center}

\subsection{Zero-crossing method}

\begin{center}
\includegraphics[width=8cm]{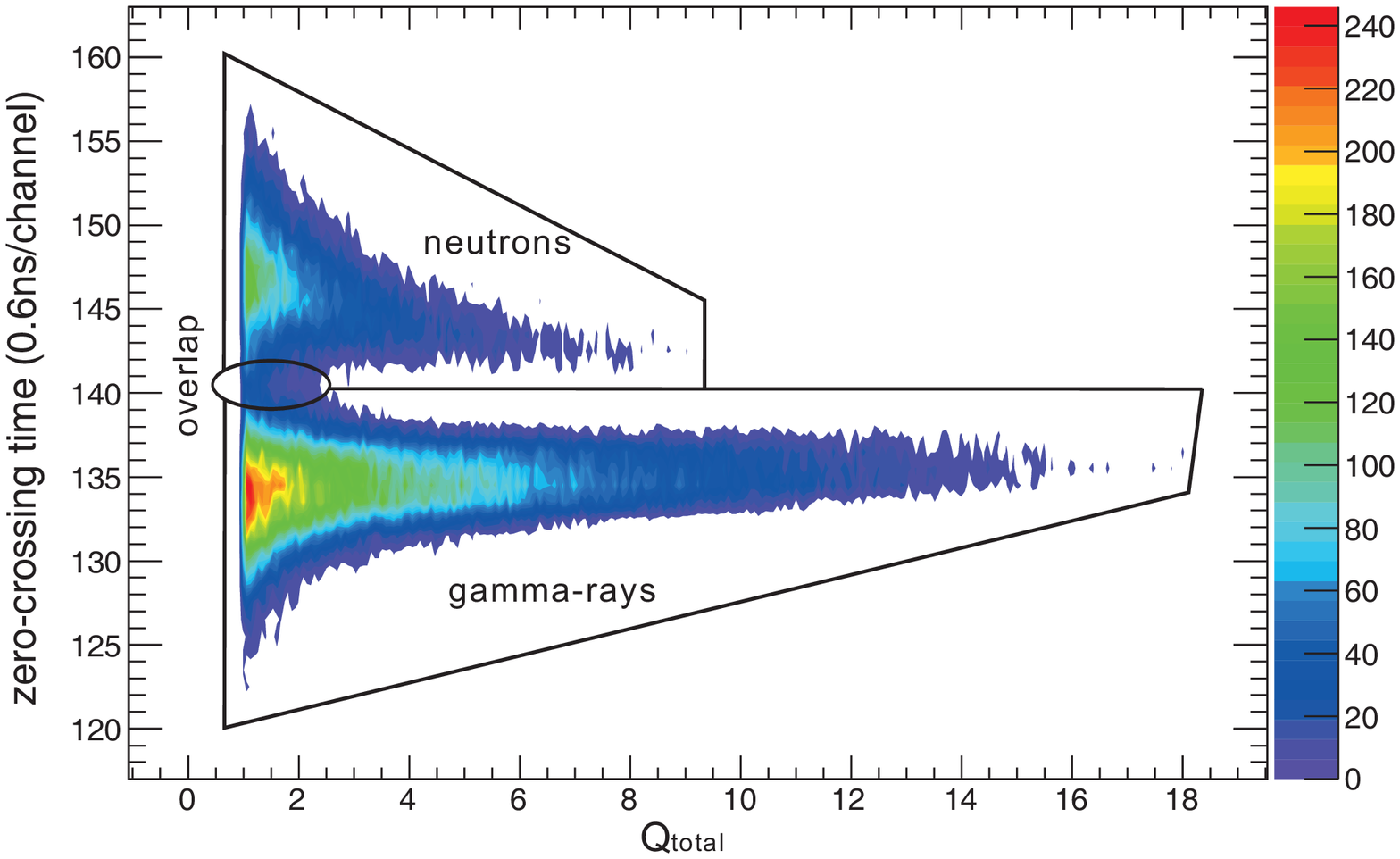}
\figcaption{\label{fig8}  2D plot of zero-crossing time versus Q$_{total}$ for $^{252}$Cf source, the energy threshold is 180 keVee. }
\end{center}

A series of FOMs were calculated at different constants for C$_1$R$_1$ and R$_2$C$_2$, and it was found that the optimum n-$\gamma$ discrimination was achieved when the differentiation time constant (C$_1$R$_1$) was 5ns and the integration time constant (R$_2$C$_2$) was 30ns.
Fig.~\ref{fig8} exhibits the zero-crossing time distribution versus Q$_{total}$ for the same experimental data set. The upper peak corresponds to neutrons because the neutron pulse crosses the baseline much later than the gamma pulse. The FOM in this case is 0.91. In order to investigate the effect of energy threshold on the quality of n-$\gamma$ separation, FOMs at 11 different energy thresholds have been calculated and shown in Fig.~\ref{fig9}. The FOMs are improved from 0.78 to 1.04 when the thresholds increase from 30 keVee to 600 keVee. This phenomenon could be explained that the influence of electronic noise decreases with larger energy threshold.

FOMs calculated using charge comparison algorithm for the same energy thresholds were also shown in Fig.~\ref{fig9} so as to compare the n-$\gamma$ separation capacity of charge comparison and zero-crossing technique. The contrastive results show that the zero-crossing method is better for energy thresholds of 80 keVee and lower. As for thresholds higher than 80keVee, the charge comparison method exhibits optimal n-$\gamma$ separation property. Therefore, the zero-crossing method is suitable for outputs lower than 80 keVee and charge comparison method is the best choice for higher equivalent electron energy. The results are similar with Ref.~\cite{lab9}, except that they investigated the comparison between charge integration and zero-crossing method using analog PSD techniques.

\begin{center}
\includegraphics[width=8cm]{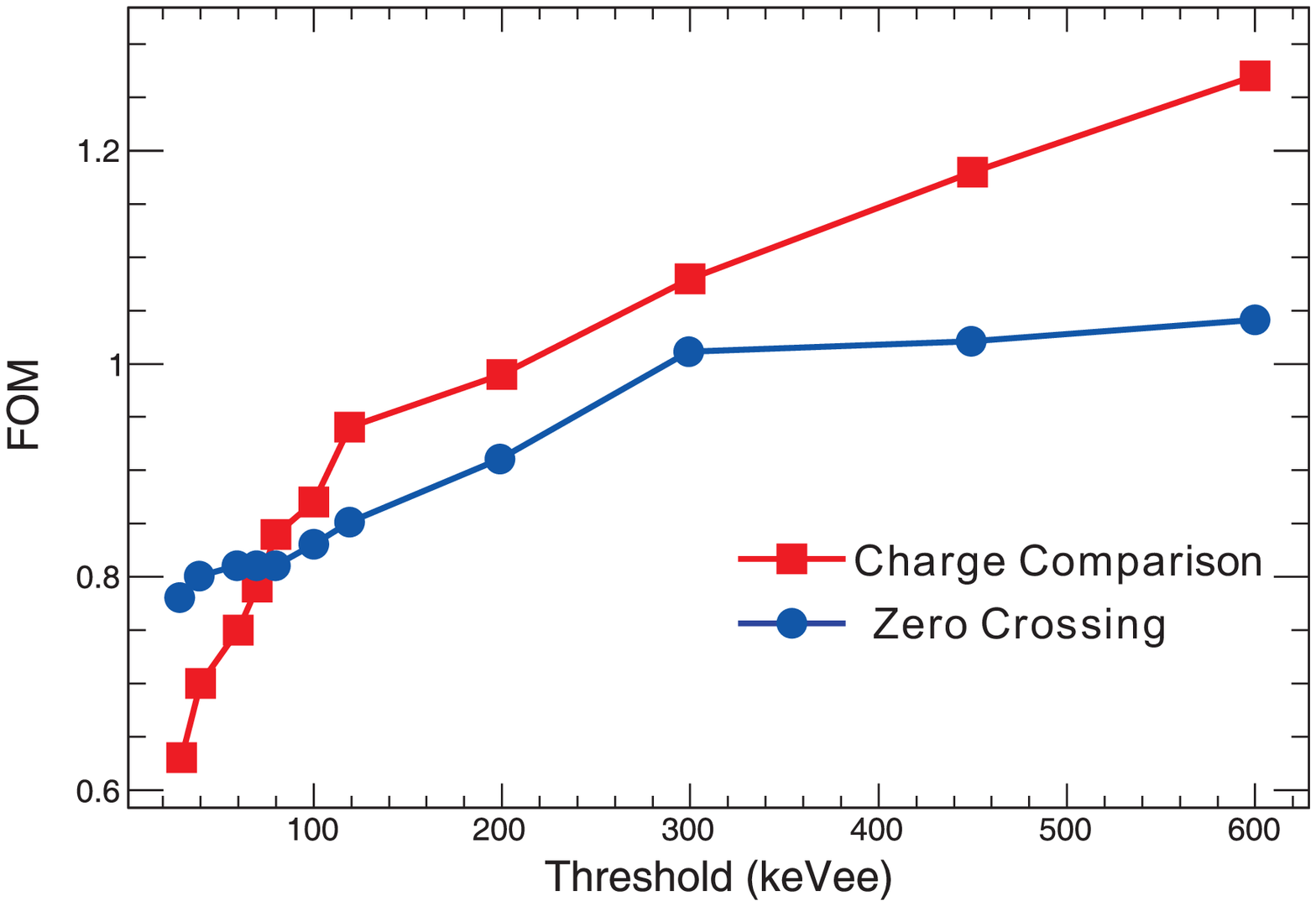}
\figcaption{\label{fig9}  Comparison of FOMs between charge integration and zero-crossing method at differen energy thresholds. }
\end{center}

\subsection{PSD for different sized detectors}

In addition, two liquid scintillation detectors mentioned in section 2.2 were tested using $^{252}$Cf source to compare the capacity of n-$\gamma$ discrimination for detectors in different size. The contrastive results calculated using charge comparison algorithm are shown in Fig.~\ref{fig10}. Significantly, the smaller liquid scintillation detector has greater FOM than the larger one at every energy threshold. This is perhaps because of longer path of photon transportation in liquid scintillation for the larger one, which distorts the time information at the tail of pulses. Consequently, for fission neutrons, the small dimension detectors are more universally used to minimize the distortion of neutron properties.

\begin{center}
\includegraphics[width=8cm]{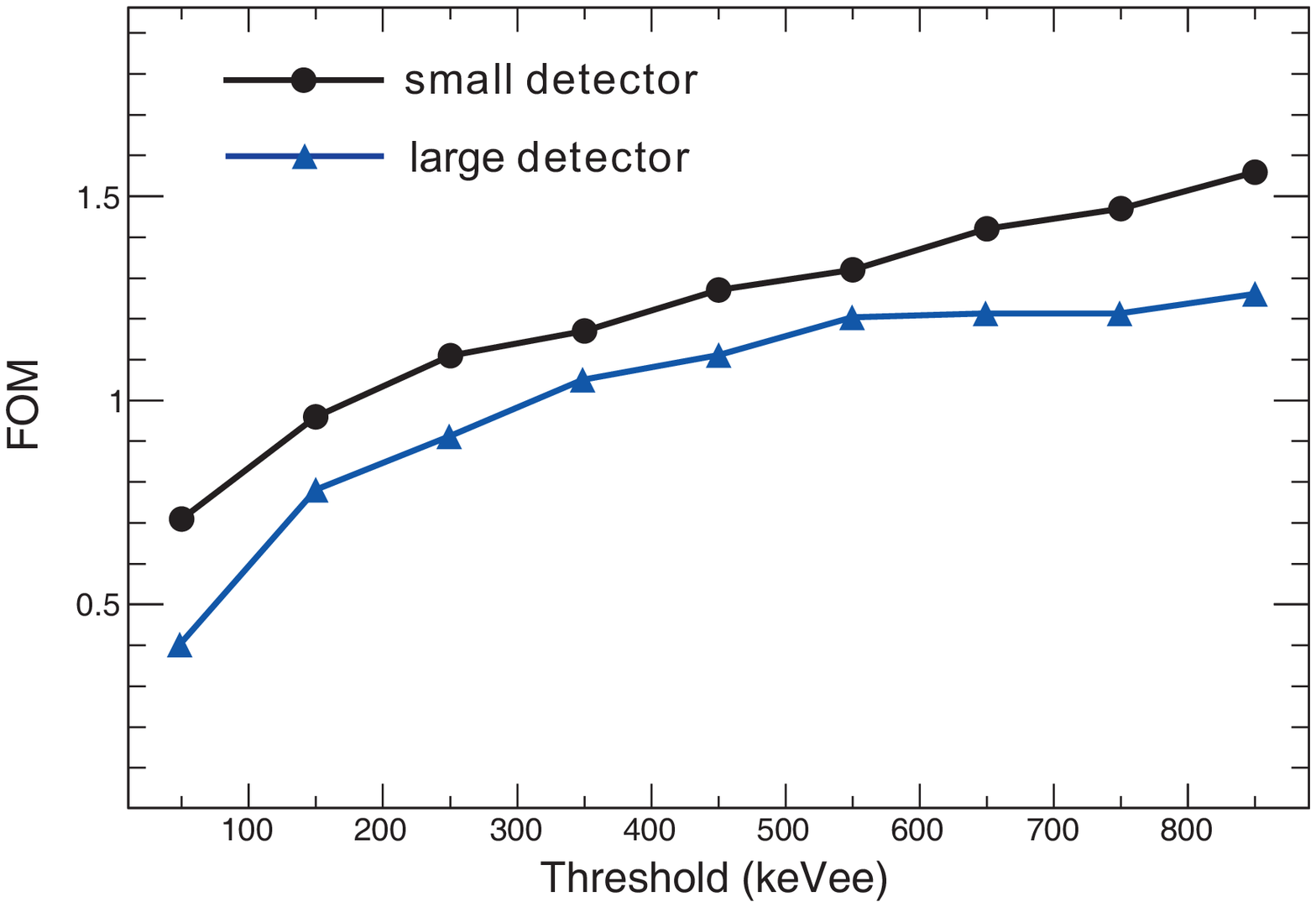}
\figcaption{\label{fig10}  Comparison of FOM for different size detectors at various energy thresholds. }
\end{center}

\section{Summary}

A digital acquisition system based on the NI-5772 adapter module for n-$\gamma$ discrimination was established and tested with a $^{252}$Cf neutron source. This system eliminates the need for QDC, TAC and delay cable ect. used in analog PSD techniques. The integration charge and zero-crossing time could be extracted through processing the digital waveforms off-line.

The energy calibration for EJ-301 liquid scintillation detector was done via $^{22}$Na, $^{137}$Cs and $^{60}$Co sources, and the gamma response function was obtained as: L=4.94*$E_e$-0.095.
Two different digital PSD algorithms were used to perform the n-$\gamma$ discrimination of EJ-301 liquid scintillation detector: (a) charge comparison method (b) zero-crossing method. Both algorithms clearly showed the power of a digital system in achieving good PSD. The digital charge comparison method presented the optimum n-$\gamma$ separation when the integration of slow components started 21ns after the pulse maximum. At energy thresholds 80 keVee and lower, the zero-crossing method gave better n-$\gamma$ discrimination. At higher energy, the charge comparison method presented better separation between neutron and gamma-ray events. In addition, as for liquid scintillation detectors in different dimension, the smaller one showed better n-$\gamma$ discrimination property for fission neutrons. The experimental results showed that such digital signal processing technique could be used for neutron monitoring of ADS and SNS in the future.

\section{Acknowledgements}

The authors thank the support of the staff who maintain the LabVIEW software package. We would also like to thank Doctor M. Nakhostin for his help on data processing.

\end{multicols}

\vspace{10mm}

\vspace{-1mm}
\centerline{\rule{80mm}{0.1pt}}
\vspace{2mm}

\begin{multicols}{2}

\end{multicols}

\clearpage

\end{CJK*}
\end{document}